\documentclass[12pt,english]{article}
\usepackage[T1]{fontenc}
\usepackage[latin1]{inputenc}
\usepackage{graphics}
\usepackage{babel}
\begin{document}
\begin{center}
{\large Comparison of cosmological models using Bayesian theory}

\vspace{.5 in}

Moncy V. John

{\it Department of Physics, St. Thomas College, Kozhencherri, 689641,
India.} 

 J. V. Narlikar

{\it Inter-University Centre for Astronomy and Astrophysics, Post
Bag 4, Ganeshkhind, Pune, 411007, India.}

\vspace{.5in}

 {\bf Abstract}
\end{center}

Using the Bayesian theory of model comparison,      a new
cosmological    model due to John and Joseph [M. V. John and K. Babu
Joseph,    Phys. Rev. D {\bf 61} 87304 (2000)] is compared with the
standard    $\Omega_{\Lambda} \neq 0$ cosmological model.  Their
analysis based    on the recent apparent magnitude-redshift data of
Type Ia    supernovas found evidence against the new model; our more 
  careful analysis finds instead that this evidence is not strong.  On
the other hand, we find that the    angular size-redshift data from
compact (milliarcsecond)    radio sources do not discriminate between
the two models.    Our analysis serves as an example of how to
compare the    relative merits of cosmological models in general,
using    the Bayesian approach.

 \medskip
\noindent PACS No(s): 98.80.Es, 02.50.-r

\newpage

\begin{center}
I. INTRODUCTION
\end{center}

In a recent publication \cite{jj}, it was argued,  by modifying an   
   earlier
ansatz by Chen and Wu \cite{chen}, that the total energy density 
$\tilde{\rho}$
for the universe should vary as $a^{-2}$ where $a$ is the scale
factor  of its expansion.
  If the total pressure is
$\tilde
{p}$, then this argument leads to $\tilde{\rho } +3\tilde {p} =0$ for 
the
universe.  This deduction was made possible by the use of some 
dimensional
considerations in line with quantum cosmology.  The reasoning is as 
follows:
Taking the comoving coordinate grid as dimensionless, we
attribute a distance dimension to the scale factor $a$. Since there is
no other fundamental energy scale available, one can 
 always
 write
$\tilde{\rho}$ as Planck density ($\rho _{pl} = c^{5}/\hbar
 G^{2}  
= 5.158
 \times 10^{93}$ g cm$^{-3}$) times a
dimensionless product of  quantities. The
variation of $\tilde {\rho }$ with  $a$ can now be
written   as

$$ \tilde{\rho} \propto \rho _{pl}\left[ \frac {l_{pl}}{a} \right]  
^{n}, $$
where $l_{pl}= (\hbar G/c^{3})^{1/2} = 1.616 \times 10^{-33}$ cm is
the    Planck
length.  It is easy to see that $n< 2$ ($n> 2$) will lead to a
negative  (positive) power of $\hbar $ appearing explicitly on the
right hand  side of the
above equation.  It was pointed out that such an $\hbar $-dependent 
total
energy density would be quite unnatural in the classical Einstein 
equation for
cosmology, much later than the Planck time.  However, the case $n=2$ is just 
  right to
survive the semi classical limit $\hbar \rightarrow 0$.  Thus it was 
argued that
if we take quantum cosmology seriously, then 
$\tilde{\rho} \propto  a^{-2}$ or
equivalently $\tilde{\rho} +3\tilde{p}=0$, for a conserved $\tilde 
{\rho}$.
Solving the Friedmann equations gives a coasting evolution for the 
universe;
i.e.,

$$ a=m\;t, $$

\noindent where $m=\sqrt {k/(\tilde{\Omega }-1 )}$ is a
proportionality  constant; $\tilde{\Omega }$ is the total density
parameter and $k=0,\pm  1$ is
the spatial curvature constant.

It shall be noted that $\tilde{\rho} +3\tilde{p}=0$ is an equation of
  state
appropriate for strings or textures and that it is unrealistic to 
consider the
present universe as string-dominated.  But in \cite{jj}, it was shown 
that this
ansatz will lead to a realistic cosmology if we consider that 
$\tilde{\rho}$
comprises of more than one component, say, ordinary matter
(relativistic or nonrelativistic) with
equation of
state $p_{m}=w\; \rho _{m}$ and a cosmological constant $\Lambda $, 
which is
time-varying.  Let $\rho _{\Lambda}$ denote the energy density arising from  
  $\Lambda
$ and $p_{\Lambda }=-\rho_{\Lambda }$ be the corresponding pressure. 
With 

$$ \tilde{\rho} = \rho_{m} + \rho _{\Lambda}, \qquad \tilde{p} =
p_{m}+p_{\Lambda} , $$

\noindent the condition $\tilde{\rho} +3\tilde{p}=0$ will give

$$ \frac{\rho_{m}}{\rho_{\Lambda}} = \frac{2}{1+3w}, $$

\noindent and this gives a realistic model for the universe.  It was 
also shown that
this  simplest cosmological model is devoid of the problems
like the horizon, flatness, monopole,
cosmological constant, size, age of the universe and the generation of 
density
perturbations on scales well above the present Hubble radius in the 
pure
classical epoch.  The solution of the cosmological constant, age and 
density
perturbation problems deserve special mention since these  are not
solvable in
an inflationary scenario.  Moreover,  the evolution of
temperature in the model is nearly the same as that in the standard big 
bang
model and  if we  assume the values $\Omega _{m}=4/3$ and $\Omega
_{\Lambda}
=2/3$, then there is no variation in the freezing temperature with the 
latter
model, and this will enable nucleosynthesis to proceed in an almost
identical  manner. It also may be noted that an almost similar model
which predicts the above values for the
density parameters was proposed  earlier  \cite{jj1},
from some more fundamental
assumptions based on  entirely different grounds.

However, it should be remarked that the  argument given above,
which leads to this cosmology, is heuristic and not based on formal
reasoning. It should be taken only as a guiding principle. Also we
note that   it has some unusual consequences like the necessity of
continuous creation of matter from vacuum energy, though it was
argued in \cite{jj,chen} that such creation will be too inaccessible
to observation.

But it was mentioned in \cite{jj} that, in spite of the those
successes in predicting  observed values, 
 the
 recent observations of the magnitudes of 42 high-redshift Type Ia 
supernovas
 \cite{perl} is a set back for the model.  A statement was explicitly 
made to
 the effect that the predictions of $\Omega _{m}$ and 
$\Omega_{\Lambda}$ for
 the present model are outside the error ellipses given in the 
$\Omega_{m}-
 \Omega_{\Lambda}$ plot in \cite{perl} and it was claimed that this
discrepancy is a 
 serious
 problem.  In this paper, we study this issue in detail to see
how strong is the evidence against this model
 when compared with the standard model with a  constant  $\Lambda
\neq 0$,      discussed in \cite{perl,riess}. Jackson and
   Dodgson
\cite{jackson1,jackson2} have examined  
 the latter model in the
light of Kellerman's
 \cite{kellerman} and Gurvits' \cite{gurvits}
compilations of angular size-redshift data for
ultracompact (milliarcsecond) radio sources.
Gurvits' compilation of such data, which are
measured  by very long-baseline interferometry
(VLBI), is claimed to have no evolution with
cosmic epoch. Several  authors (for e.g.,
\cite{jvn})  have made use of these data to test
their cosmological models. In the present paper,
we also analyze the Gurvits' data to test the new
model.  Using the Bayesian theory of  statistics,
we compare the new model discussed above with the
standard model  with a non-zero cosmological constant,  using
both the apparent magnitude-redshift data and the
angular size-redshift data. It is found that
there is no strong evidence against the new model
when the apparent magnitude-redshift data are
considered. This is contradictory to the
statement made in \cite{jj}. The angular
size-redshift  data, on the other hand, are found
to provide equal preference to the 
standard model and   the new
one.

The remainder of this paper takes the new theory as given and
compares it with  other standard cosmological models. The
analysis shall be viewed as an example of using Bayesian theory to
test the relative merits of cosmological models, a method which is
claimed to have many positive features when compared to indirect
arguments using parameter estimates. As such, the technique described
here has wider applicability than just to the comparison of two
cosmological models.

The paper is organized as follows. In Section II, we discuss the
Bayesian  theory of  model comparison for the general case.
Section III discusses comparison of the two models  using  apparent
magnitude-redshift data and in Section IV, we compare the models with
the angular size-redshift data. Section V comprises discussion of
the results.

\begin{center}
II. BAYESIAN THEORY OF MODEL COMPARISON

\end{center}

The Bayesian theory of statistics \cite{loredo,rome} is historically 
the
original approach to statistics, developed by great mathematicians
like Gauss, Bayes, Laplace, Bernoulli etc., and has several
advantages over the  currently used long-run relative frequency
(frequentist) approach to statistics, especially in problems like
those in astrophysics, where the notion of  a statistical ensemble is 
highly contrived.  The frequentist definition of probability can only
describe the probability of a true random variable, which
can take on various values throughout an ensemble or a series of 
repeated
experiments.  In astrophysical and similar problems, ensembles and repeated
experiments are rarely possible and we speak about
 the  probability of a
hypothesis, which can only be either true or
 false,  and hence
is not a random variable.   The Bayesian theory
will   help 
assign probabilities for such hypotheses by considering the (often 
incomplete)
data available with us.  For example, Laplace used Bayesian theory to 
estimate
the masses of planets from astronomical data, and to quantify the 
uncertainty
of the masses due to observational errors \cite{jaynes}. In fact,
this 
 theory finds
 application in all those problems where one can only have a 
numerical
encoding of one's state of knowledge.

In the Bayesian theory of model comparison, it is common to report 
model
probabilities via odds, the ratios of probabilities of the models.  The
posterior (i.e., after consideration of the data) odds for the model 
$M_{i}$
over $M_{j}$ are

$$
O_{ij} = \frac{p(M_{i}|D,I)}{p(M_{j}|D,I)}, 
$$ 
 
 \noindent where $p(M_{i}|D,I)$ refers to the posterior probability for 
the
model $M_{i}$, given the data $D$ and assuming that any other 
information $I$
regarding the models under consideration is true.  Using Bayes's
theorem, one
can write the above equation as

\begin{equation}
 O_{ij} = \frac{p(M_{i}|I){\cal L}(M_{i})}{p(M_{j}|I){\cal
L}(M_{j})}, \label{eq:oij1} 
\end{equation}

\noindent where $p(M_{i}|I)$ is called the prior probability; i.e., any
probability assigned to the model $M_{i}$ before consideration of the 
data, but assuming the information $I$ to be true.
When  $I$ does not give any preference to one model over
the
other, these prior probabilities are equal so that

\begin{equation}
 O_{ij} = \frac{{\cal L}(M_{i})}{{\cal L}(M_{j})} \equiv B_{ij}.
\label{eq:oij2} 
\end{equation}

\noindent $B_{ij}$ is called the Bayes factor.  ${\cal L} (M _{i})$
denotes the probability $p(D|M_i )$ to obtain the data $D$ if the
model $M_i$ is the true one and is referred to as the likelihood for
the model $M_i$. The models under consideration
will usually have one or more free parameters
(like the density parameters $\Omega_m $, $\Omega
_{\Lambda }$ etc. in the case of cosmological
models), which we denote as $\alpha$, $\beta $,
.. . ${\cal L} (M _{i})$ can be evaluated for 
models with one parameter as

\begin{equation} {\cal L}(M_{i}) \equiv p(D|M_{i}) =\int d\alpha \;
p(\alpha|M_{i}){\cal L}_{i}(\alpha), \label{eq:likem} \end{equation}

\noindent where $p(\alpha|M_{i})$ is the prior probability for the
parameter
$\alpha$, assuming the model $M_{i}$ to be true.  ${\cal
L}_{i}(\alpha)$ is the
likelihood for $\alpha$ in the model and is usually taken to have the
form

\begin{equation}
{\cal L}_i (\alpha ) \equiv \exp [-\chi ^2 (\alpha )/2].
\label{eq:litheta}
 \end{equation}

\noindent where 

\begin{equation}
\chi ^2 = \sum_k \left( \frac {\hat {A }_k -A_k (\alpha ) }{\sigma_k
}\right) ^2
 \end{equation}

\noindent is the $\chi^2$ statistic.  Here $\hat {A}_k $ are the
measured
values of the observable $A$, $A_k (\alpha )$ are its expected values
(from
theory) and $\sigma _k$ are the uncertainties in the measurements of 
the
observable.

Generalization to the case of more than one parameter is straight
forward. As a specific case, consider a model $M_i $ with two
parameters, $\alpha $ and $\beta $, having flat prior probabilities;
i.e., we assume to have no prior information regarding $\alpha $ and 
$\beta $ except that they lie in some range [$\alpha $, $\alpha
+\Delta \alpha $] and [$\beta $, $\beta +\Delta \beta $], 
respectively. Then  $p(\alpha |M_i ) = 1/\Delta \alpha $,
$p(\beta |M_i ) = 1/\Delta \beta $ and hence

\begin{equation}
{\cal L}(M_i )= \frac {1}{\Delta \alpha } \frac {1}{\Delta \beta }
\int _{\alpha }^{\alpha +\Delta \alpha }d\alpha
\int _{\beta }^{\beta +\Delta \beta }d\beta \exp [-\chi^2 (\alpha
,\beta )/2] .  \label{eq:Lalphabeta}
\end{equation}
It is instructive to rewrite this equation as

$$
{\cal L}(M_i)=\frac{1}{\Delta \alpha }\int _{\alpha}^{\alpha + \Delta
\alpha} d\alpha {\cal L}_i (\alpha ).
$$
In this case,

$$
 {\cal L}_i (\alpha )=\frac{1}{\Delta \beta }\int _{\beta}^{\beta +
\Delta \beta} d\beta \exp [-\chi ^2 (\alpha ,\beta )/2]
$$
is called the marginal likelihood for the parameter $\alpha $.

\begin{center}
A. Interpretation of the Bayes factor

\end{center}

The interpretation of the Bayes factor $B_{ij}$, which is given by
(\ref{eq:oij2}) and which evaluates the relative merits of model
$M_i$   over
model $M_j$, 
is as follows \cite{drell}:  If
$1<B_{ij} <3$, there is an evidence against $M_j$ when compared with 
$M_i$, but it
is not worth more than a bare mention.  If $3<B_{ij} <20$, the
evidence   against
$M_j$ is definite but not strong.  For $20<B_{ij} <150$, this evidence
is strong and 
 for $B_{ij}
>150$, it is very strong.

 \begin{center}
III. COMPARISON USING REDSHIFT-MAGNITUDE DATA

\end{center}

For an FRW model which contains matter and a cosmological constant, the
likelihood for these parameters, i.e., ${\cal L}_{i}(\Omega _{m},\Omega
_{\Lambda })$ can be assigned using the redshift-apparent magnitude 
data in the
following manner \cite{drell}. Before consideration of the data, let
us agree that $\Omega_m $ lies somewhere in the range $0<\Omega_m <3$,
 $\Omega_{\Lambda }$ in the range $-3<\Omega_{\Lambda }<3$ and
accept this as the only prior information $I$.
  Let $\hat {\mu}_{k}$ be the observed
best-fit
distance modulus for the supernova number $k$, $s_{k}$ its uncertainty 
and
$\hat{z}_{k}$ is the  cosmological redshift, with $w_{k}$ its 
uncertainty.
We can write the expression for $\chi^2 $ as

\begin{equation}
\chi ^2 = \sum_k \left( \frac {\hat {\mu }_k -\mu_k  }{\sigma_k
}\right) ^2 .
 \end{equation}

\noindent Here,

\begin{equation} 
\hat{\mu}_{k} =\mu_{k}+n_{k} = g_{k}-\eta +n_{k}, \label{eq:muk}
\end{equation}

\noindent with

$$
\mu_{k} \equiv g_{k} -\eta =5\; \log \left[ \frac {D_L(z;\Omega_m,
\Omega_{\Lambda },H_0 )}{1\hbox{Mpc}}\right] +25
$$

\noindent being the redshift-apparent magnitude relation. The
luminosity  distance is
$D_L(z;\Omega_m, \Omega_{\Lambda },H_0) = cH_0^{-1} d_L (z;\Omega_m , 
\Omega
_{\Lambda })$, where $c$ is the velocity of light, $H_0$ is the Hubble 
constant
at the present epoch and $d_L$ is the dimensionless luminosity 
distance.
$g_{k}=g(\hat {z}_{k})$ is the part of $\mu_k$ which depends
implicitly  on
$\Omega _{m}$ and $\Omega _{\Lambda}$ and $\eta $ is its 
$H_{0}$-dependent
part. The latter quantity can be written as $\eta \equiv 5 \log
(h/c_2 )-25 $  where $H_0 = h\times 100$ km s$^{-1}$ Mpc$^{-1}$ and
$c_2 $ is the speed of light in units of 100 km s$^{-1}$.
 The probability distribution for the value $n_{k}$ in Equation
(\ref{eq:muk})  is assumed to be a
zero-mean Gaussian with standard deviation $\sigma _{k}$, where
$\sigma  _{k}^{2} = s_{k}^{2}+
[\mu^{\prime}(\hat{z}_{k})]^{2}w_{k}^{2}$, in the  absence
of systematic or evolutionary effects.

One can evaluate ${\cal L}(\Omega_m,
\Omega _{\Lambda }, \eta )$ in a manner similar to that in 
(\ref{eq:litheta}),
where $\chi^2$ now is a function of the three parameters
$\Omega_m$, $\Omega _{\Lambda }$ and
$\eta $.  The likelihood for $\Omega_m$ and $\Omega _{\Lambda }$,
denoted as
${\cal L}(\Omega_m, \Omega _{\Lambda })$ can be obtained by 
the technique of marginalising over
$\eta $, if one assumes a flat prior probability for $\eta $ in some 
appropriate range.

 To do this, we define $s^{-1}=\sqrt {\sum _{k}(1/\sigma
_{k}^{2})}$ where $s$ is the posterior uncertainty for $\eta $ and
let
 $1/\Delta \eta$ a flat prior probability be assigned to $\eta $.
 (These 
   being the
same for all models, will get canceled when evaluating probability 
ratios.)
Using these definitions, the marginal likelihood (defined at the
end of Sec. II) for the density
 parameters is 

\begin{equation}
 {\cal L}(\Omega _{m}, \Omega_{\Lambda}) = \frac {1
}{\Delta \eta}\int d\eta e^{-\chi^2 /2}. \label{eq:likeo0} 
\end{equation}

\noindent Evaluating this integral analytically \cite{drell}, one
assigns a  likelihood for
the parameters $\Omega _{m}$ and $\Omega _{\Lambda }$ in any one
model   as

\begin{equation} {\cal L}(\Omega _{m}, \Omega_{\Lambda}) = \frac {s 
\sqrt
{2\pi}}{\Delta \eta}e^{-q/2}, \label{eq:likeo} \end{equation}

\noindent where

\begin{equation} 
q(\Omega _{m}, \Omega _{\Lambda }) = \sum _{k} \frac {(\hat
{\mu}_{k} -g_{k}+\hat {\eta})^{2}}{\sigma _{k}^{2}}, \label{eq:q} 
\end{equation}

\noindent  is of the form of a $\chi ^{2}$-statistic, with 
$\hat
{\eta}$ the best fit (most probable) value of $\eta$, given $\Omega 
_{m}$ and
$\Omega_{\Lambda }$. The latter can be computed as \cite{drell}

\begin{equation} 
\hat{\eta } (\Omega _{m}, \Omega _{\Lambda }) = s^2 \sum _{k} \frac 
{(g_k -\hat
{\mu}_{k} )^{2}}{\sigma _{k}^{2}}. \label{eq:etahat} 
\end{equation}

 Now, we compare the model in
\cite{perl,riess} (model $M_1$, having parameters
$\Omega _{m}$,  $\Omega_{\Lambda}$ and $\eta $)
with the new model discussed in Sec.   I (model
$M_2$, having only the parameters $\Omega _m$ and
$\eta$). The
Bayes factor $B_{12}$ can be written with the help of Eq.  
(\ref{eq:oij2}) and
 Eq.  (\ref{eq:likem})  as

\begin{equation} B_{12}= \frac {{\cal L}(M_{1})}{{\cal L}(M_{2})}
=\frac {\int
d\Omega_{m}\;
\int d\Omega _{\Lambda } p(\Omega_{m},\Omega_{\Lambda }|M_{1}){\cal 
L}_{1}(\Omega_{m},\Omega
_{\Lambda })}{\int
d\Omega_{m}\; p(\Omega _{m}|M_{2}){\cal L}_{2}(\Omega_{m})}. 
\label{eq:o121} \end{equation} 

 With the information $I$ at  hand, one can assign flat
prior probabilities  $p(\Omega_{m},\Omega_{\Lambda }|M_{1})=1/18$
and $p(\Omega _{m}|M_{2})=1/3$ .  Using Eqs. (\ref{eq:Lalphabeta})
and    (\ref{eq:likeo}) we can write the above as

\begin{equation} B_{12}=
\frac {\int
_{-3}^{3 } d\Omega _{\Lambda } \int _{0}^{3 } d\Omega _{m}
\exp
[-q_{1}(\Omega _{m},\Omega _{\Lambda })/2]}{6\int _{0}^{3 } d\Omega
_{m} \exp [-q_{2}(\Omega  _{m})/2]}.  \label{eq:o122}
\end{equation}

Our first step in the evaluation of $B_{12}$ is to find $q$ given in
Eq.  (\ref{eq:q}), for both the models. For Model 1,  we have to use

$$ g(z)=5\log \{ (1+z)\; |\Omega_k |^{-1/2}
\hbox{sinn}[|\Omega_k |^{1/2} I(z)]\} , $$

\noindent where    $\Omega_k =1-\Omega_m -\Omega_{\Lambda }$ and 
$\hbox{sinn}(x) =\sin x$ for $\Omega_m + \Omega_{\Lambda }>1$,
$\hbox{sinn}(x) =\sinh x$ for $\Omega_m + \Omega_{\Lambda }<1$  and
$\hbox{sinn}(x) = x$ for $\Omega_m + \Omega_{\Lambda }=1$. Also

$$ I(z) = \int _{0}^{z} [(1+z^{\prime })^{2}(1+\Omega
_{m}z^{\prime})-z^{\prime}(2+z^{\prime} )(\Omega _{\Lambda
})]^{-1/2}\;  dz^{\prime}. $$

\noindent For Model 2, the function $g(z)$  can be written as

$$ g(z)=5 \log \{ m(1+z)\hbox{sinn}[\frac {1}{m}\ln (1+z)] \} , $$

\noindent where
$m=\sqrt {2k/(3\Omega _{m} -2)}$ for the nonrelativistic era and 
$\hbox{sinn}(x) =\sin x$ for $\Omega_m >2/3$,
$\hbox{sinn}(x) =\sinh x$ for $\Omega_m <2/3$  and
$\hbox{sinn}(x) = x$ for $\Omega_m =2/3$.

Using these expressions,  Eq.  (\ref{eq:o122}) is numerically
evaluated  to obtain $B_{12}= 3.1$.  (In
this
calculation, we have used the data corresponding to the Fit C in 
\cite{perl},
which involve 54 supernovas.) As per the interpretation of
 $B_{ij}$ given in Sec. II.A,  the above is an evidence against Model
2,  but it is only barely definite; the discrepancy is not a
"serious problem" as had been stated   in \cite{jj}.

\begin{center}

IV. COMPARISON USING ANGULAR SIZE-REDSHIFT DATA

\end{center}

For this purpose, we use the Gurvits' data
and divide the  sample which contains 256
sources into 16 redshift bins, as done by Jackson 
and Dodgson and shown in their
Fig. 1 \cite{jackson2}.  For Model 1, we use the
expression for angular size

\begin{eqnarray}
\Delta \theta =\frac {d}{d_A } & \equiv &\frac  {d} { (1+z)^{-1}
\left( k/\Omega_k \right)^{1/2}
\frac{c}{H_0 }\hbox{sinn}[|\Omega_k |^{1/2}I^{\prime}(z)] }  \nonumber
\\
 & = & \frac{dH_0 }{c}
\frac{(1+z)}{\left(k/\Omega_k \right)^{1/2}
\hbox{sinn} [|\Omega_k |^{1/2}I^{\prime}(z)]},
\end{eqnarray}

\noindent where

\begin{equation}
I^{\prime}(z)=\int_1^{1+z} \frac{dx}{x\left(\Omega_k +\Omega_m x+
\frac {\Omega_{\Lambda }}{x^2}\right)^{1/2}}.
\end{equation}

\noindent Here $d$ is the linear dimension of an object, $d_A $ is
the angular  size distance and $\Omega_k $ and $\hbox{sinn}(x)$ are
defined as in the case of Model 1 in the last section. Similarly for
Model 2, we have

\begin{equation}
\Delta \theta =\frac {d}{d_A }= \frac{dH_0 }{c}
\frac{(1+z)}{ m\; \hbox{sinn} [\frac{1}{m}\ln (1+z)]},
\end{equation}

\noindent where $m$ and $\hbox{sinn}(x)$ are defined as in the
earlier
 case of  Model 2. For the purpose of comparison, we only 
need to
 combine the unknown parameters $d$ and $H_0$ to form a
single
 parameter $p\equiv dH_0 /c$.  Thus Model 1 has three
parameters $p$,
 $\Omega_m $ and $\Omega_{\Lambda }$ whereas Model 2
has only the
 parameters $p$ and $\Omega_m $. As in the
previous case, we accept $0<\Omega_m <3$ and 
$-3<\Omega_{\Lambda }<3$  as the prior
information $I$. With these ranges of values of
$\Omega_m$ and $\Omega_{\Lambda }$, $p$ is found
to give significantly low values of $\chi^2 $ only
for the  range $0.1<p<1$ in both the models, $p$
being given in units of milliarcseconds.  The formal
expressions to be used are

\begin{equation}
\chi ^2 = \sum_k \left( \frac {\hat {\Delta \theta }_k -\Delta
\theta_k  }{\sigma_k  }\right) ^2
 \end{equation}

\noindent and

\begin{eqnarray}
B_{12} =\frac {{\cal L}(M_1)}{{\cal L}(M_2 )} &
= &\frac  
{
\frac {1}{\Delta p} \frac{1}{\Delta \Omega_m}
\frac {1}{\Delta \Omega_{\Lambda }} \int dp\int d\Omega_m \int
d\Omega_{\Lambda }  \exp [-\chi_{1}^{2}(p,\Omega_m ,\Omega_{\Lambda
})]} {
\frac {1}{\Delta p} \frac{1}{\Delta \Omega_m}
 \int dp\int d\Omega_m
\exp [-\chi_{2}^{2}(p,\Omega_m )]} \nonumber \\ 
& = &\frac  {\int_{0.1} ^1 dp
\int_0 ^3 d\Omega_m  \int_{-3} ^{3} d\Omega
_{\lambda } \exp [-\chi_1 ^2 /2]}{6 \int_{0.1} ^1 dp \int_0 ^3
d\Omega_m \exp [-\chi_2 ^2 /2]}. \label{eq:b12last}
\end{eqnarray}

The result obtained is $B_{12} \approx 1 $. This
may be interpreted  as providing equal preference
to both models.

\begin{center}

V. DISCUSSION

\end{center}

While evaluating the Bayes factors using both kinds of data,
 we have assumed that our prior
information $I$ regarding the density parameters is  $0<\Omega
_{m}<3$ and  $-3<\Omega_{\Lambda }<3$. The range of values of
$\Omega_{\Lambda }$ considered in \cite{perl} is
$-1.5<\Omega_{\Lambda }<3$ and in
\cite{jackson2} it is $-4<\Omega_{\Lambda }<1$. 
Even if we modify the range of this parameter in
our analysis to some reasonable extent, the main
conclusions of the paper will remain unaltered.
For example, if we
accept $0<\Omega_m <3$ and $-1.5<\Omega_{\Lambda} <1.5$ as some prior
information $I^{\prime}$, the Bayes factors in each case become 3.8
and 0.8, in place of 3.1 and 1, respectively. Instead, if we choose
$I^{\prime}$ as $0<\Omega_m <3$ and $-6<\Omega_{\Lambda}<6$, the
corresponding values are 1.55 and 1.4, respectively. These do not
change our conclusions very much in the light of the discriminatory
inequalities mentioned in Sec. II.A. 

\begin{figure}[h] 
\centering{\resizebox {0.9 \textwidth} {0.9 \textheight } 
{\includegraphics {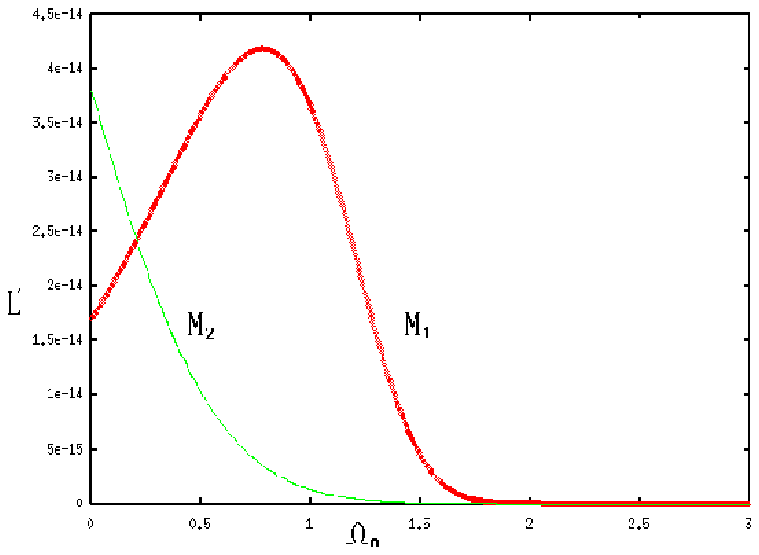}} 
\caption{$L^{\prime}$ vs $\Omega_m $  for both models, using the
apparent  magnitude-redshift  data for Type Ia supernova.  The
 curves  $M_1$ and $M_2$ correspond to the marginal likelihoods for
$\Omega_m$  for the standard $\Omega_{\Lambda } \neq 0$ model 
and the new model, respectively (apart from some multiplicative
constants, which cancel on taking ratios).  }  \label{fig:fig1}}   
\end{figure}

\begin{figure}[h] 
\centering{\resizebox {0.9 \textwidth} {0.9 \textheight } 
{\includegraphics {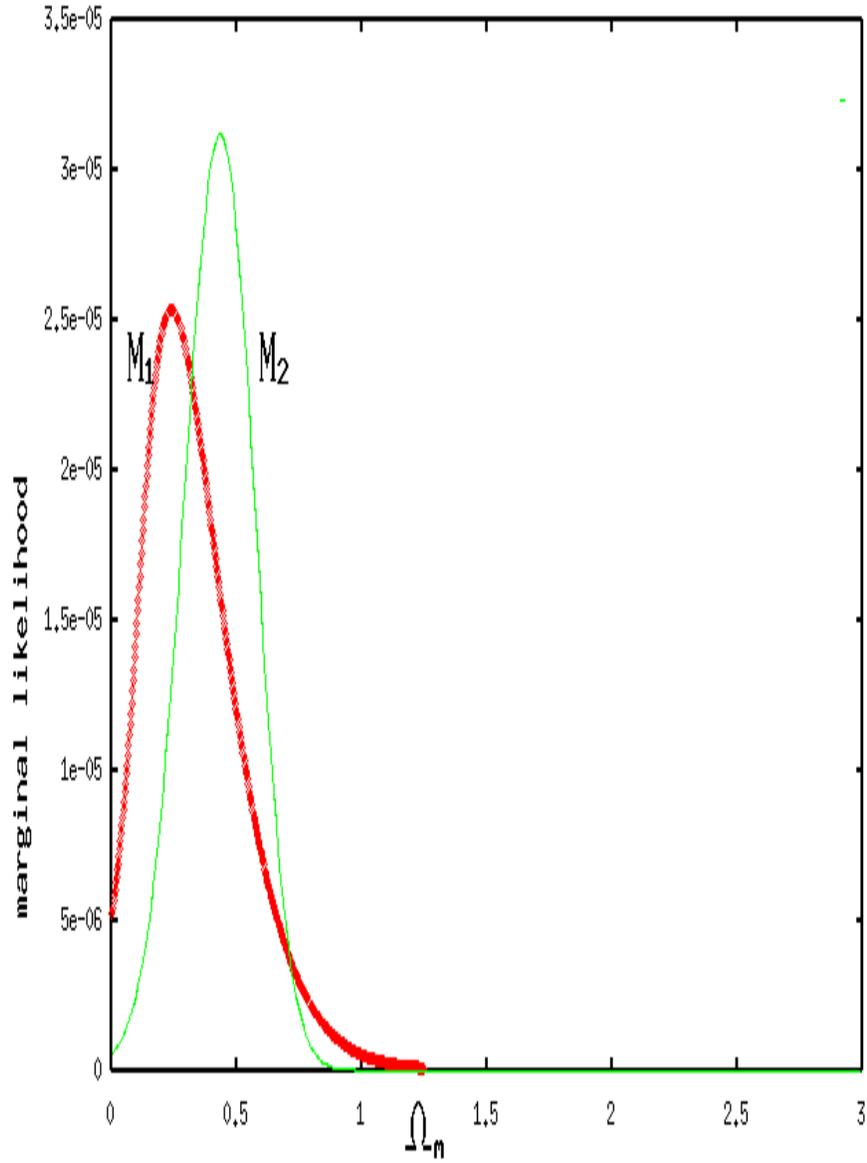}} 
\caption{Marginal likelihood vs $\Omega_m $  for both models, using
the angular size-redshift  data.  The  curves $M_1$ and $M_2$ 
correspond to the marginal likelihoods for $\Omega_m$  for the
standard $\Omega_{\Lambda } \neq 0$ model    and
the new model, respectively.  }  \label{fig:fig2}}    \end{figure}

In order to get an intuitive feeling why  the standard
($M_1$) and new ($M_2$) models have  comparable  likelihoods, consider
Figs. \ref{fig:fig1} and \ref{fig:fig2}. Fig. \ref{fig:fig1} is for
the apparent magnitude-redshift data and plots the quantities 
$L^{\prime }= \frac{1}{6}\int _{-3}^{3}d\Omega _{\Lambda }\exp
[-q_1(\Omega_m ,\Omega_{\Lambda })/2]$ (curve labeled $M_1$) and
$L^{\prime }=\exp [-q_2(\Omega _m )/2] $ (curve labeled $M_2$)
against $\Omega _m$. From the definition of marginal likelihood given
at the end of Sec. II and from Eqs. (9)-(14), it can be seen that
these two curves correspond to the marginal likelihoods for the
parameter $\Omega_m$ in models $M_1$ and $M_2$, respectively (apart
from some  multiplicative constants, which cancel on taking ratios).
Similarly, Fig. \ref{fig:fig2}, which is  for the angular
size-redshift data, plots ${\cal L}=\frac{1}{6\times 0.9}\int
_{0.1}^{1} dp \int_{-3}^{3} d\Omega_{\Lambda }\exp [-\chi_1^2/2]$
(curve $M_1$) and ${\cal L}=\frac{1}{0.9}\int _{0.1}^{1}dp \exp
[-\chi_2^2/2]$ (curve $M_2$) against $\Omega_m$. Eq.
(\ref{eq:b12last}) allows us to interpret these terms as the marginal
likelihoods for $\Omega_m$ in models $M_1$ and $M_2$, respectively.
In fact,  these curves
rigorously show the integrands one must integrate over $\Omega_m$ to
get the Bayes factors. Using the apparent magnitude-redshift data,  a
lower value of  value of  $q$ (which is a modified $\chi ^2 $
statistic) is obtained for model $M_1$ whereas for angular size-red
shift data, lower  $\chi^2$ is claimed by  model $M_2$. However, the
areas under the curves are comparable in both cases and this shows why
the Bayes factors are also comparable. This is one of the  strong
points of the Bayesian method, in contrast to  frequentist 
goodness of fit tests, which  consider only the best fit 
parameter values  for comparing models \cite{loredo}. 
 
These figures, however, show some feature that is disturbing for the
new model. Figs. \ref{fig:fig1} and \ref{fig:fig2} indicate best fit
values of $\Omega_m =0$ and $\Omega_m= 0.42$, respectively, for this
model. In both cases it appears to rule out the value $\Omega_m=4/3$
that is needed to meet the constraints on nucleosynthesis, a
condition which had been stated in the introduction.  Though, as
mentioned above, Bayesian model comparison does not hinge upon the
best fit values in evaluating relative merits of models, one would
desire to have an agreement between predicted and observed
parameter values. A natural option in such cases would be to
compare the models by adjusting the prior regarding the parameters so
that  any additional information is accounted for. But we have not
attempted this in our analysis.

The constant $\Omega_{\Lambda} \neq 0$ model we considered has one
parameter in excess of the new model in both cases. It should be kept
in mind that in the Bayesian method, simpler models with less number
of parameters are often favored unless the data are truly difficult
to account for with such models. Bayes's factors thus implement a
kind of automatic and objective Occam's razor. In this context, it is
interesting to check how the new model fares when compared with  flat
(inflationary) models where $\Omega_m +\Omega_{\Lambda }=1$, by which
condition the number of parameters of model $M_1$ are reduced by one.
This makes the two models at par with each other, with regard to the
number of parameters. We have calculated the Bayes factor between
this flat model $M_1$ and the new model $M_2$, using the apparent
magnitude-redshift data and the result is $B_{12}=5.0$. This appears
to be a slightly more definite evidence against the new model than
the corresponding result obtained in Sec. III ($B_{12}=3.1$).
(However, inflationary models with a constant $\Lambda $-term suffer
from the `graceful exit problem' for $\Lambda $; i.e., in order to
explain how $\Lambda $ manages to change from its GUT magnitude to
$\approx 10^{-126}$ of its initial value, some extreme fine tuning
would be required \cite{wein}).  On the other hand, a comparison of
$\Omega_m +\Omega_{\Lambda }=1$ model with the new model using
angular size-redshift data gives a value for the Bayes factor
$B_{21}=15$, which shows that this data is  difficult to account
with the flat inflationary models  than with the new one. The results
we obtained, while using the information $I$, are summarized in Table
I.

 When compared to the frequentist goodness of fit
test of models, which judges
the relative merits of the models using the lowest
value of $\chi^2 $ (even when it is obtained by some fine tuning
or by having more parameters), the present
approach has the advantage that it evaluates the
overall performance of the models under
consideration. 
The Bayesian method is thus a very powerful tool
of model comparison and it   is
high time  that the method is used to evaluate the plausibility of
cosmological
models cropping up in the literature.  It is true that since we have 
only one
universe, one can only resort to model making and then to comparing 
their predictions with
observations.  Again, since we cannot experiment with the universe,
it   is not
meaningful to use the frequentist approach.  We believe that the only
meaningful way
 is   to use
the Bayesian approach in such cases.  Here we have made a comparison 
between the  model in \cite{perl,riess} with the
new model in \cite{jj}.   It
deserves to be stressed that the recent
  apparent magnitude-redshift
observations  on Type Ia supernovas do not pose a "serious problem"
to  the new model, as had been claimed in \cite{jj}. The angular
size-redshift data, on the other hand, do not discriminate between the
general
 $\Omega_{\Lambda }\neq 0$ model and the new model 
and they provide definite but not strong evidence against standard
flat ($\Omega _m + \Omega _{\Lambda }=1$) model when compared to
the new one.

  Here it is essential to point out that Bayesian inference
summaries the weight of evidence by the full posterior odds and not
just by the Bayes factor. Throughout our analysis above, we have
assumed that the only prior information with us is either $I$ (stated
in the beginning of Sec. III) or $I^{\prime}$ (stated in the
beginning of Sec. V), which helps to make the posterior odds equal to
the Bayes factor. However, when the Bayes factor is near unity, the
prior odds $p(M_i\mid I)/p(M_j\mid I)$ in Eq. (\ref{eq:oij1}) become
very important. The standard $\Omega_{\Lambda} \neq 0$ model and the 
standard flat (inflationary) models are plagued by the large number
of cosmological problems (as mentioned in Sec. I) and the new model
has the heuristic nature of its derivation and the problem with
nucleosynthesis, setting (subjective) prior odds against each of
them. In the context of having obtained comparable values for the
Bayes factor, the Bayesian model comparison forces us to conclude, in
a similar tone as in \cite{drell}, that the existing apparent
magnitude or angular size-redshift data alone are not very
discriminating about these cosmological models. It is also worth
remarking  here that the Bayesian theory  tells us how to adjust our
plausibility assessments when our state of knowledge regarding an
hypothesis changes through the acquisition of new data \cite{loredo}.
Concerning future observations, one would have to say that if the
supernova test is extended to higher redshifts and if the astronomers
are sure about the standard candle hypothesis, then the theories can
be tested for such new data using Bayesian model comparison, using
what we have now obtained  as the prior odds.   In this context, it
also deserves serious consideration to extend the analysis made here 
to other cosmological data, like that of cosmic microwave background
radiation and primordial nucleosynthesis. Hopefully, further analysis
and future observations may help to give more decisive answers on
these questions.

{\bf Acknowledgements}

We are thankful to Professor K. Babu Joseph, Ninan Sajith Philip and
Dr. R. G. Vishwakarma 
 for  valuable discussions.
Also the  help in performing the computations
rendered by Ninan Sajith Philip  and Mathew V.
Samuel are  acknowledged with thanks.


\newpage

\begin{center} 
 
Table I.  
 
\end{center} 
 
\begin{tabular} {l l l l l} \hline \hline
 
Data  & Model $M_1$ & Model $M_2$ & Bayes factor& Interpretation \\
\hline
$m-z$ & Standard  & New model &
$B_{12}=3.1$ & Slightly definite but \\  
 &$\Omega_{\Lambda }\neq 0$ model & & & not strong evidence \\
 & & & & against the new model \\
$m-z$ & Standard flat & New model & $B_{12}=5$ & Definite but \\
   & $\Omega_{\Lambda}\neq 0$ model &  &  & not strong evidence \\
 & & & & against the new model \\

$\theta - z$ & Standard  & New model &
$B_{12}=1$ & Both models are \\
 &$\Omega_{\Lambda }\neq 0$ model & & & equally favored \\ 
$\theta - z$ & Standard flat & New model & $B_{21}=15$ & Definite but
\\
& $\Omega_{\Lambda}\neq 0$ model &  &  & not strong evidence \\
& & & & against the flat model \\ \hline \hline
 \end{tabular}

\end{document}